\documentclass[twocolumn,preprintnumbers,amsmath,amssymb,amsfonts,prb,floatfix]{revtex4}

\usepackage{graphicx} 
\usepackage{dcolumn} 
\usepackage{bm} 

\begin{document}

\title{Green's Function Technique for Studying Electron Flow in 2D Mesoscopic Samples}  

\author{G.~Metalidis}
  \email{georgo@mpi-halle.de}
\author{P.~Bruno}
  \email{bruno@mpi-halle.de}
\affiliation{%
Max-Planck-Institut f\"{u}r Mikrostrukturphysik, Weinberg 2,
D-06120 Halle, Germany} \homepage{http://www.mpi-halle.de}
\date{\today}

\begin{abstract}
In a recent series of scanning probe experiments, it became
possible to visualize local electron flow in a two-dimensional
electron gas. In this paper, a Green's function technique is
presented that enables efficient calculation of the quantity
measured in such experiments. Efficient means that the
computational effort scales like $M^3 N$ ($M$ is the width of the
tight-binding lattice used, and $N$ is its length), which is a
factor $MN$ better than the standard recursive technique for the
same problem. Moreover, within our numerical framework it is also
possible to calculate (with the same computational effort $M^3 N$)
the local density of states, the electron density, and the current
distribution in the sample, which are not accessible with the
standard recursive method. Furthermore, an imaging method is
discussed where the scanning tip can be used to measure the local
chemical potential. The numerical technique is used to study
electron flow through a quantum point contact. All features seen
in experiments on this system are reproduced and a new
interference effect is observed resulting from the crossing of
coherent beams of electron flow.
\end{abstract}

\maketitle

\section{Introduction}
Electronic transport properties in mesoscopic systems have gained
a lot of attention in the last two decades (for a review see e.g.
Refs.~\onlinecite{Datta, Ferry}). A lot of effort already went
into the study of global transport quantities (mainly
conductance), and a range of interesting phenomena emerged: e.g.
quantized conductance in a point contact~\cite{vanWees}, quantum
Hall effect~\cite{vonKlitzing}, and universal conductance
fluctuations~\cite{Lee} to name a few.

Recently, scanning probe methods have offered the interesting
possibility of obtaining local information on electron transport,
not accessible with ordinary global measurements. For example, by
using the tip of an STM as a localized scatterer of electrons and
measuring the conductance change of the sample as a function of
tip position, it has been possible to obtain a spatial map of
electron flow in a two-dimensional electron
gas~\cite{TopinkaScience, TopinkaNature, TopReview}. Such
visualizations of current flow are interesting both from an
experimental and a theoretical point of view.

The results obtained in these experiments were mainly interpreted
with the help of electron or flux density calculations. Here, a
numerical technique is developed that allows to calculate in an
efficient way the quantity that is actually measured, which is a
conductance difference as a function of the tip position. The
standard recursive Green's function method used in
Ref.~\onlinecite{HeZhuWang} to obtain the same result is not very
efficient because one has to start over the complete conductance
calculation for every position of the tip. The numerical effort
then scales like $M^4 N^2$ in the number of operations, where $M$
is the width and $N$ is the length of the system ($N\gg1$), and
one therefore is limited to small systems. On the other hand, our
method scales like $M^3N$ for the same problem. Moreover, with our
method it is straightforward to obtain other relevant information
like the local density of states and the electron and current
density distributions which can then be compared with the
experimental observables. All these quantities can be calculated
with the same computational effort scaling like $M^3 N$, whereas
they are unavailable with the standard technique.

The imaging technique with the tip as a scatterer will not always
give the expected results in the regime of high magnetic fields.
To obtain more information about electron flow in this high field
range, a measuring method is proposed where the STM tip is used as
a local voltage probe. This method can also be treated within the
proposed numerical framework, with a computational effort scaling
again as $M^3 N$.

The paper is subdivided as follows. In the next section, the
imaging method used in Refs.~\onlinecite{TopinkaScience,
TopinkaNature} is briefly reviewed and the voltage probe method is
described. In section~\ref{sysmodel}, we discuss how the
considered systems can be described within a tight-binding model.
Subsequently, the numerical method is presented in some detail in
section~\ref{numericalmethod} and in the appendices. Finally, the
method is applied to study electron flow through a quantum point
contact.

\section{Imaging Methods}

\subsection{Tip as a Local Scatterer} \label{scatterer}
The experiments in Refs.~\onlinecite{TopinkaScience,
TopinkaNature} take place as follows: current is passed via two
Ohmic contacts through a two-dimensional electron gas, while
simultaneously a scanning probe tip is moved across the sample.
The local electrostatic potential that results from a negative
voltage on the tip can function as a scattering center for
electrons in the device. If the tip is placed over a region where
a lot of electrons are flowing, the conductance of the sample will
decrease considerably because of enhanced backscattering due to
the tip, whereas in an area of low electron flow, the conductance
decrease will be small. As such, by mapping the conductance
decrease of the sample to the tip position, one gets a spatial
profile of electron flow.

The relevant quantity for this imaging method is thus the position
dependent conductance difference:
\begin{equation} \label{eqdeltag}
\Delta g (x,y) = g_0 - g_t(x,y),
\end{equation}
where $g_t(x,y)$  is the conductance with the tip positioned on a
point $(x,y)$ and $g_0$ is the conductance without the tip.
Conductances can be calculated within the Landauer-B\"{u}ttiker
formalism, which relates conductance to the electron transmission
coefficient $T$ between the contacts on opposite sides of the
sample:
\begin{equation} \label{eqLandauer}
g = \frac{2e^2}{h} T.
\end{equation}
Please note that since $g$ is a two-terminal conductance, it is
symmetric with respect to reversal of the direction of an applied
magnetic field: $g(+\mathbf{B}) =
g(-\mathbf{B})$~\cite{Buettiker}, and therefore the map of
electron flow obtained with this scattering method has the same
symmetry.

\subsection{Tip as a Voltage Probe} \label{voltprobe}
The imaging method in the previous section can give very nice
visualizations of the electron flow, as proven by the experimental
results of electron transport through a quantum point
contact~\cite{TopinkaScience, TopinkaNature}. One of the
interesting subjects to study with this imaging method would be
the quantum Hall effect, because even after intensive theoretical
and experimental effort since its discovery~\cite{vonKlitzing},
some of the details of electron transport in this regime remain
unclear. Unfortunately, the appearance of edge states in the high
field regime leads to the suppression of
backscattering~\cite{Butt}, so that the method described above
will not yield the expected results in the quantum Hall limit.

However, a picture of this edge-state transport can be obtained by
investigating the local chemical potential in the sample. This can
be done using the STM tip as a voltage probe; the STM tip voltage
equilibrates itself to the local chemical potential when electrons
are allowed to tunnel into and out of the tip. Experimentally,
this technique has already been used to probe the potential
distribution at metal-insulator-metal interfaces~\cite{Muralt} and
at grain boundaries~\cite{Kirtley}. Gramespacher et al. have
studied this measurement mode theoretically, and found Green's
function expressions for the sample-tip transmission coefficients
which generalize the Bardeen formulas~\cite{Gramespacher1}. For
weak coupling between the probe and the sample, these transmission
coefficients can be expressed in terms of so-called partial
densities of states~\cite{Gramespacher2}.

In the following, we will however start again from the standard
Landauer-B\"{u}ttiker formalism, but now one has three contacts:
the left and right contact through which a fixed current is
passed, and the STM tip that measures a voltage. For this 3-lead
structure at temperature $T=0$, one obtains the following
expression for the voltage measured on the tip~\cite{Datta}:
\begin{equation} \label{eqvolttip}
V_{\text{tip}} - V_L = \frac{T_{\text{tip},L}}{T_{\text{tip},L} +
T_{\text{tip},R}} \, (V_R - V_L),
\end{equation}
with $V_p$ the voltage on lead $p$. Our numerical method thus
should be able to calculate the transmission coefficients
$T_{\text{tip},L(R)}$ from the left and right contact to the STM
tip, as a function of the position of the tip.

It should be indicated that this imaging mode corresponds to
making a three-terminal measurement. This means that the spatial
map one obtains will not be invariant under the reversal of the
direction of an applied magnetic field. The only symmetry
relations that hold are: $T_{pq}(+\mathbf{B}) =
T_{qp}(-\mathbf{B})$~\cite{Buettiker}. Therefore, it should be
clear that the information obtained when the tip is used as a
voltage probe is different from that when the tip is used as a
local scatterer, and as such both imaging methods will contribute
differently to our understanding of electron flow in 2D systems.

\subsection{Charge and Current Density}
One can (at least in principal) experimentally observe the total
current density in the sample, by measuring the magnetic field it
induces. Another physically observable quantity is the total
charge density, due to the electrostatic field it generates. Both
of these quantities can also be calculated within the numerical
framework presented in the next sections.

Since we have access to the electron density, it would also be
possible to include self-consistently the effect of the applied
voltage and the resultant current flow on the transport
properties. However, this will not be pursued further in the
current paper.

\section{System Modelling} \label{sysmodel}

The system we will consider is a two-dimensional device connected
to two semi-infinite leads extending in the x-direction. By
discretizing the Schr\"{o}dinger equation, one obtains the
standard nearest neighbor tight-binding model for the system
Hamiltonian:
\begin{widetext}
\begin{equation} \label{eqtb}
H = \sum_{n=-\infty}^{\infty} \sum_{m=1}^{M} \epsilon_{mn}
|m,n\rangle \langle m,n| + \sum_{n=-\infty}^{\infty}
\sum_{m=1}^{M-1} \left( t_{m,n}^{x} |m,n+1\rangle \langle m,n| +
t_{m,n}^{y} |m+1,n\rangle \langle m,n | + h.c. \right).
\end{equation}
\end{widetext}
where $(m,n)$ label the sites on the lattice, and $\epsilon_{mn}$
are the on-site energies. Note that the sites are not meant to
represent atoms as in the usual tight-binding model; rather they
may represent a region containing many atoms, but this region
should be small with respect to physically relevant quantities
such as the Fermi wavelength. The quantities $t_{m,n}^{x}$ and
$t_{m,n}^{y}$ in Eq.~(\ref{eqtb}) give the hopping amplitude in
the horizontal, respectively vertical direction. A square lattice
is assumed, so in the absence of a magnetic field one has:
\begin{equation}
t_{m,n}^{x} = t_{m,n}^{y} = -t = -\frac{\hbar^2}{2 m^{\ast} a^2},
\end{equation}
with $a$ the lattice spacing, and $m^{\ast}$ the effective mass of
the electron.

The leads extend from $n=-\infty..0$ for the left lead, and
$n=N+1,..,\infty$ for the right one. The device itself is
comprised by the other $M \times N$ sites. In the leads, only
homogeneous fields perpendicular to the 2D device will be
considered, while both homogeneous and inhomogeneous fields can be
treated in the central device. The fields are described by Peierls
substitution~\cite{Peierls}, which changes the hopping parameters
as:
\begin{equation}
t_{mn}^{x(y)} = -t \, \mathrm{e}^{-\mathrm{i} \, e / \hbar \int
\mathbf{A} \centerdot \mathrm{d} \mathbf{l}},
\end{equation}
where $\int \mathbf{A} \centerdot \mathrm{d} \mathbf{l}$ is the
integral of the vector potential along the hopping path.

\section{Numerical Method} \label{numericalmethod}

\subsection{Introduction} \label{nummethodintroduction}
\begin{figure}
\includegraphics{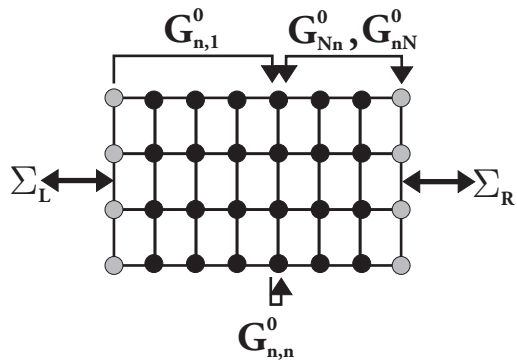}
\caption{\label{Fig1} View on the Green's functions necessary to
calculate the relevant physical quantities in our problem. The
first and last column of the tight binding lattice are colored
grey to indicate that they are influenced by the self-energies
$\Sigma_L$ and $\Sigma_R$ of the leads.}
\end{figure}

In sections~\ref{scatterer} and~\ref{voltprobe}, it was shown that
the basic quantities that need to be calculated in order to
describe the scanning probe experiments are transmission
coefficients from one lead to another. In the next sections, it
will be shown that these transmission coefficients, and as a
matter of fact all other quantities we wish to calculate (density
of states, current density distribution,...) can be expressed in
terms of the Green's function matrices $G_{n1}^{0}$, $G_{nn}^{0}$,
$G_{nN}^{0}$ and $G_{Nn}^{0}$ (for all possible values of $n$).
These $M \times M$ matrices $G_{l l'}^{0}$ are defined as:
\begin{equation}
\langle m | G_{ll'}^{0} | m' \rangle = \langle m,l | G^0 | m',
l'\rangle,
\end{equation}
where
\begin{equation} \label{eqGreendef2}
G^0 = \frac{1}{E + \mathrm{i} \eta - H^d - \Sigma_L - \Sigma_R},
\end{equation}
is the Green's function of the device \textit{without} the
influence of the tip. In this expression, $H^d$ is the Hamiltonian
of the central device disconnected from the leads, while
$\Sigma_L$ and $\Sigma_R$ are self-energies of the left
respectively right lead. The Green's functions $G_{ll'}^{0}$ are
thus submatrices of $G^{0}$ connecting points from column $l'$ to
those of column $l$ of the lattice (see also Fig.~\ref{Fig1}).

The method that is used to calculate these Green's functions will
be presented in Appendix~\ref{AppendixA}, the main result being
that they can be obtained with a number of operations that scales
like $M^3 N$. As for the self-energies of the leads, they can be
calculated in several ways: in the absence of a magnetic field
they are known analytically~\cite{Datta}, while in the case of a
homogeneous magnetic field one needs to resort to numerical
methods, for instance based on the solution of the so-called
Harper's equation~\cite{Guan}. However, we have chosen for another
method, originally developed for the calculation of surface
electronic structure, that is described in full detail in
Refs.~\onlinecite{LopezSancho, Turek}.

\subsection{Local Scatterer Method} \label{locscat}
In our calculations, the scattering potential created by the STM
tip is modelled by a delta function located on a site $(m,n)$, so
it adds a repulsive contribution $V^{\text{tip}} = v | m,n \rangle
\langle m,n |$ to the on-site energies in the
Hamiltonian~(\ref{eqtb}). Like explained in
section~\ref{scatterer}, one has to calculate a conductance
difference:
\begin{equation} \label{eqtransw}
\Delta g (m,n) = \frac{2 \mathrm{e}^2}{h} [ T^0 - T(m,n) ],
\end{equation}
for every tip position $(m,n)$ in order to obtain the spatial map
of electron flow in the sample. The transmittance $T(m,n)$ between
the left and right lead can be calculated by~\cite{Datta}:
\begin{equation} \label{eqtransmission2}
T(m,n) =  \textrm{Tr} \, [\Gamma_R \, G_{N1} \, \Gamma_L \,
G_{N1}^{\dag}],
\end{equation}
with $\Gamma_{L(R)}$ related to the retarded self-energies
$\Sigma_{L(R)}$ of the left (right) lead as:
\begin{equation} \label{eqGamma}
\Gamma_{L(R)} = \mathrm{i} \, (\Sigma_{L(R)} -
\Sigma_{L(R)}^{\dag}),
\end{equation}
and $G_{N1}$ is the Green's function between columns $1$ and $N$.
The matrix $G_{N1}$ includes the effect of the STM tip, and is
therefore different from $G_{N1}^{0}$, mentioned in the
introduction.

In Ref.~\onlinecite{HeZhuWang}, one uses the standard recursive
Green's function method for calculating $G_{N1}$: first, the
system is divided into its separate columns (by making the hopping
matrices between them zero), and the repulsive potential
$V^{\text{tip}}$ is added on a certain site. Subsequently the
columns are attached one by one making use of Dyson's equation.
For this attachment procedure, $N-1$ inversions of an $M \times M$
matrix are needed to obtain $G_{N1}$ for a single tip position.
Since this procedure has to be started over and over again for
every single position of the tip, the total number of inversions
needed to calculate $G_{N1}$ for all positions scales as $M N^2$
(in the limit of large N), which translates into a total
computational effort scaling like $M^4 N^2$ (computational effort
to invert an $M \times M$ matrix scales as $M^3$).

However, supposed that we have access to the Green's functions
$G_{n1}^{0}$, $G_{Nn}^{0}$ and $G_{nn}^{0}$ (see
Appendix~\ref{AppendixA}), we are able to do things more
efficient, by including the effect of the tip with Dyson's
equation:
\begin{equation} \label{eqDyson}
G = G^{0} + G^{0} \, V \, G,
\end{equation}
where $V$ is the potential introduced by the tip, $V =
V^{\text{tip}}$. Projecting (\ref{eqDyson}) between columns $1$
and $N$, one obtains (it is assumed that the tip is located at
lattice site $(m,n)$):
\begin{widetext}
\begin{equation}
G_{N1} = G_{N1}^{0} + G_{Nn}^{0} \, V_{nn}^{\text{tip}}\, (1 -
G_{nn}^{0} \, V_{nn}^{\text{tip}})^{-1} \, G_{n1}^{0}.
\label{eqDysonperturb}
\end{equation}
\end{widetext}
Since $V^{\text{tip}}$ has only one non-zero element, the
inversion $(1 - G_{nn}^{0} \, V_{nn}^{\text{tip}})^{-1}$ will boil
down to the inversion of a scalar. This means that no extra matrix
inversions are needed to find $G_{N1}$ for an arbitrary lattice
position of the tip, once we have the $G_{Nn}^{0}$, $G_{nn}^{0}$
and $G_{n1}^{0}$ for all $n$! In Appendix~\ref{AppendixA}, we will
show a way of calculating these functions with a number of matrix
inversions that scales linear with $N$, so that the computational
effort for calculating $G_{N1}$ for all tip locations scales as
$M^3 N$.

At first sight, it might seem that this efficiency is decreased
for the complete calculation because one has to evaluate the trace
in Eq.~(\ref{eqtransmission2}) for every tip position, which
involves products of $M \times M$ matrices. The computational
effort for doing such a product scales like $M^3$, and since there
are $MN$ lattice sites, the total effort would scale as $M^4 N$.
However, we have a better way of evaluating this trace, scaling as
$M^3 N$, so that we do not loose our efficiency. Technical details
will be described in Appendix~\ref{AppendixB}.

It now is clear that our method, scaling like $M^3 N$, is more
efficient than the standard recursive technique, which scales like
$M^4 N^2$ for the same problem.

\subsection{Voltage Probe Method}
In this case, the STM tip will be modelled by a one-dimensional
semi-infinite lead, attached to the central device at position
$(m,n)$ (the lead can be thought to extend in a direction
perpendicular to the 2D sample). The voltage on the tip can be
written as a function of the transmission coefficients
$T_{\text{tip}, L}$ and $T_{\text{tip}, R}$ between the STM tip
and the left and right leads (Eq.~(\ref{eqvolttip})). These
transmittances, with the tip positioned over site $(m,n)$, can be
expressed as:
\begin{subequations} \label{eqtransmissionvolt}
\begin{eqnarray}
T_{\text{tip}, L} &=&  \text{Tr} \, [\Gamma_{\text{tip}} \, G_{n1}
\, \Gamma_L \, G_{n1}^{\dag}], \\
T_{\text{tip}, R} &=&  \text{Tr} \, [\Gamma_{\text{tip}} \, G_{nN}
\, \Gamma_R \, G_{nN}^{\dag}],
\end{eqnarray}
\end{subequations}
where $\Gamma_{\text{tip}} = \mathrm{i} \, (\Sigma_{\text{tip}} -
\Sigma_{\text{tip}}^{\dag})$. Since the lead modelling the tip is
one-dimensional, one does not have to take care of magnetic field
effects in this lead, and the self-energy $\Sigma_{\text{tip}}$ of
the tip is known analytically~\cite{Datta}:
\begin{equation}
\Sigma_{\text{tip}} = -t \, \mathrm{e}^{\mathrm{i}
(Acos(1-E/(2t)))}.
\end{equation}

The standard recursive Green's function technique only gives
access to $G_{N1}$, so it cannot be used to obtain the Green's
functions $G_{n1}$ and $G_{nN}$ in
expression~(\ref{eqtransmissionvolt}). But once again, we can use
Dyson's equation~(\ref{eqDyson}) to relate these Green's functions
to Green's functions of the device without the tip (calculated in
Appendix~\ref{AppendixA}):
\begin{subequations}
\begin{eqnarray}
G_{n1} &=& (1-G_{nn}^{0} \, V_{nn}^{\text{tip}})^{-1} \, G_{n1}^{0}, \\
G_{nN} &=& (1-G_{nn}^{0} \, V_{nn}^{\text{tip}})^{-1} \,
G_{nN}^{0},
\end{eqnarray}
\end{subequations}
where the tip influence leads to an imaginary potential
$V^{\text{tip}} = \Sigma_{\text{tip}} \, |m n \rangle \langle mn
|$.

Again, the inversion $(1-G_{nn}^{0} \, V_{nn}^{\text{tip}})^{-1}$
will reduce to the inversion of a scalar, because $V^{\text{tip}}$
has only one non-zero element. In this case, calculation of the
traces in Eqs.~(\ref{eqtransmissionvolt}) is also not
computationally expensive since $\Gamma_{\text{tip}}$ has only one
non-zero element. Therefore the total computational effort scales
like $M^3 N$, needed for the calculation of the Green's functions
without the influence of the tip (see Appendix~\ref{AppendixA}).

\subsection{LDOS - Electron Density}
In order to compare the visualizations of electron flow obtained
by the two imaging methods above with the theory of electron
transport, it can be useful to calculate the local density of
states (LDOS) and the electron density distribution in the sample,
of course without any STM tip over the sample. For the LDOS, the
standard expression is~\cite{Datta}:
\begin{equation}
\rho(E, m,n) = -\frac{1}{\pi} \, \text{Im} \, \langle m |
G_{nn}^{0}(E) | m \rangle.
\end{equation}

Having calculated the linear density of states, it is also easy to
obtain the electron density in the sample by integrating over
energy.

\subsection{Current Density Distribution} \label{curdensdist}
When a magnetic field is present, persistent currents are flowing
through the device, even in the absence of an applied bias. In a
recent paper by Cresti et al.~\cite{Cresti}, an expression for
this equilibrium current is derived from the Keldysh formalism.
Adapted to our notation, the expression for the particle current
at temperature $T=0$ flowing from one node to a neighboring node
reads (remember that $m$ labels the rows of the lattice, $n$ the
columns):
\begin{widetext}
\begin{subequations} \label{eqcurrenteq}
\begin{eqnarray}
I_{(m,n-1)\rightarrow (m,n)}^{\text{eq}} &=& \frac{2 e}{\hbar}
\int_{0}^{E_F} \frac{\textrm{d}E}{2 \pi} \  2 \, \text{Re}\,
\langle m | \left[ G_{nn}^{0} \, \Sigma_{n}^{\text{left}} -
\Sigma_{n}^{\text{left}} \, G_{nn}^{0} \right] |m
\rangle, \\
I_{(m,n)\rightarrow(m+1,n)}^{\text{eq}} &=& \frac{2 e}{\hbar}
\int_{0}^{E_F} \frac{\textrm{d}E}{2 \pi} \  2 \, \text{Re} \,
\langle m + 1| \left[ t_{m,n}^y \, (G_{nn}^{0} - G_{nn}^{0 \,
\dag}) \right] | m \rangle.
\end{eqnarray}
\end{subequations}
\end{widetext}
Here, $E_F$ is the Fermi energy of the device, and $e$ is the
negative electronic charge. We have also introduced:
\begin{equation}
\Sigma_{n}^{\text{left}} = V_{n, n-1} \, G_{n-1, n-1}^{0,L}
\, V_{n-1,n}, \\
\end{equation}
where $V_{n,n-1}$ describes the hopping between columns $n-1$ and
$n$, and the Green's functions $G_{nn}^{0,L}$ are defined in
Fig.~\ref{Fig8} of Appendix~\ref{AppendixA}.

It is clear, by taking the trace over the row indices $m$, that
the total current flowing through every single column is equal to
zero, so as expected in equilibrium there will be no net current
through the leads. Also, when no magnetic field is present, all
Green's functions in the Eqs.~(\ref{eqcurrenteq}) are symmetric so
that the equilibrium current density in this case vanishes like it
should.

In the non-equilibrium situation, there are two contributions to
the current density. A magnetic field gives rise to persistent
currents, while the applied bias leads to a transport current.
Since the persistent currents are anti-symmetric with respect to
the direction of the field, we can define a transport current as
the symmetric part of the total current density distribution. This
transport current is gauge invariant, and corresponds to a
physically relevant (and measurable) quantity. In the linear
response regime, it is given by (see also
Ref.~\onlinecite{Cresti}):
\begin{widetext}
\begin{subequations} \label{eqcurrentneq}
\begin{eqnarray}
I_{(m,n-1)\rightarrow (m,n)} &=& \frac{-2
e^2 V}{2 \pi \hbar} \ \left\{ A(E)|_{+B} + A(E)|_{-B} \right\}, \quad \text{with} \nonumber \\
A(E) &=& \ 2 \, \text{Im} \, \langle m | \left[ G_{nn}^{0} \,
\Gamma_{n}^{\text{right}} \, G_{nn}^{0 \, \dag} \,
(\Sigma_{n}^{\text{left}})^{\dag} \right] | m \rangle \\
I_{(m,n)\rightarrow(m+1,n)} &=& \frac{-2 e^2 V}{2 \pi \hbar} \
\left\{C(E)|_{+B} + C(E)|_{-B} \right\}, \quad \text{with} \nonumber \\
C(E) &=& \ 2 \, \text{Im} \, \langle m+1 | \left[ t_{m,n}^y \,
G_{nn}^{0} \, \Gamma_{n}^{\text{right}} \, G_{nn}^{0 \, \dag}
\right] | m \rangle
\end{eqnarray}
\end{subequations}
\end{widetext}
where $V = V_L - V_R$ is the potential difference between the
leads, and:
\begin{subequations}
\begin{eqnarray}
\Sigma_{n}^{\text{left}} &=& V_{n, n-1} \, G_{n-1, n-1}^{0,L}
\, V_{n-1,n}, \\
\Sigma_{n}^{\text{right}} &=& V_{n, n+1} \, G_{n+1, n+1}^{0,R} \,
V_{n+1,n}, \\
\Gamma_{n}^{\text{right}} &=& \mathrm{i} \,
[\Sigma_{n}^{\text{right}} - (\Sigma_{n}^{\text{right}})^{\dag} ].
\end{eqnarray}
\end{subequations}
The Green's functions $G_{nn}^{0, L(R)}$ in these expressions are
defined in Fig.~\ref{Fig8} of Appendix~\ref{AppendixA}.

\section{Applications}
In the previous sections, we have obtained quite a lot of
quantities that can give relevant information about the flow of
electrons in a two-dimensional electron gas. To show the power of
our method, we discuss two physical systems in this section.

\subsection{Single Quantum Point Contact}
The first system that we consider is the one that is used in the
experiment of Topinka et al.~\cite{TopinkaNature}. It consists of
a 2DEG with a quantum point contact in the middle. The point
contact is modelled by a potential of the form:
\begin{equation} \label{eqQPCpotential}
W \textrm{exp}^{-x^2/\xi^2} y^2.
\end{equation}
We have taken a lattice of 1001 by 351 sites, with a lattice
parameter of $a=6.2 \, \textrm{nm}$, which corresponds to a
hopping parameter $t=14.5 \, \textrm{meV}$ (the effective mass of
the electron is taken to be that for electrons in GaAs: $m^{\star}
= 0.068 \, m$). The parameters of the potential are chosen as:
$W=0.56 \, t$ and $\xi=10 \, a$. The Fermi energy is put equal to
$E_F = 1.1 \, t = 16 \, \textrm{meV}$ (corresponding to a
wavelength $\lambda_F = 6 \, a$), which is on the first
conductance plateau of the point contact. For the calculations
where the tip functions as a local scatterer, the strength of the
scattering potential is chosen to be $v = 8t$. Disorder in the
system is modelled by a plane of impurities above the 2DEG, where
the repulsive potential from a single impurity is taken to vary
with distance $r$ as $1/r^3$, which is characteristic for the
screened potential in a 2DEG from a point charge~\cite{Davies}.
The concentration of impurities is fixed at 1\% of the total
number of lattice sites. The impurity lattice is located at a
distance $6 \, a$ above the 2DEG. Within the Born approximation,
the mean free path of the potential we use is estimated to have a
value of $4 \cdot 10^3 \, a$ which is much longer than the system
size so that we are in the ballistic regime. The mobility
corresponding to these parameters is of the order $\mu \approx 2
\cdot 10^6 \, \textrm{cm}^2/\textrm{Vs}$.
\begin{figure}
\includegraphics[width=8cm]{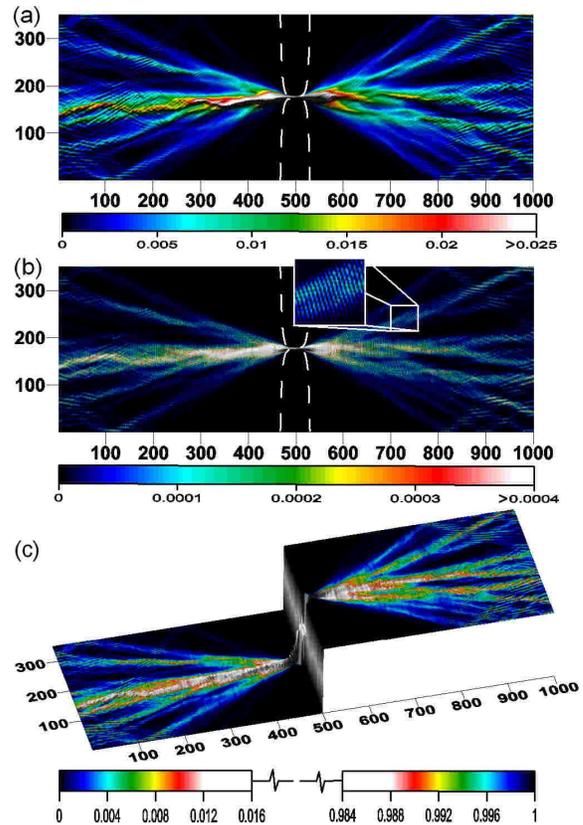}
\caption{\label{Fig2} Maps of electron flow through a quantum
point contact. Lengths on the axes are given in units of the
lattice parameter $a$. Current density distribution~(a) in units
of $2e^2V/(h a)$, STM conductance map~(b) (conductance is measured
in units of $2e^2/h$) and STM volt probe map~(c). The voltage on
the left lead $V_L = 0$, and voltage is measured in units of
$V_R$. The Fermi energy contour is depicted as a dotted white
line. There is a clear correspondence between the different
images.}
\end{figure}

In Fig.~\ref{Fig2}a, the calculated current density is seen to
exhibit a branching behavior that was also apparent in the
experiment in Ref.~\onlinecite{TopinkaNature}. The branched flow
is also present in the map of the conductance difference versus
the tip position, when the tip is used as a local scatterer
(Fig.~\ref{Fig2}b). It is clear that there is a direct
correspondence between the current density calculation and the
conductance map. Therefore one can conclude that the experiment in
Refs.~\onlinecite{TopinkaScience, TopinkaNature} really probes the
current distribution in the sample. Also visible in the
conductance difference map are the interference fringes spaced by
half the Fermi wavelength (see the inset in Fig.~\ref{Fig2}b),
which were explained as resulting from scattering between the
point contact and the STM tip~\cite{TopinkaScience}. The map of
the local chemical potential, as measured by the STM voltage probe
(Fig.~\ref{Fig2}c) in this case gives similar information as the
previous plots: on the left the current flow appears as regions
with increased voltage compared to that of the left lead ($V_L$ is
put equal to zero). This corresponds to a decreased chemical
potential due to a deficit of electrons resulting from the
non-equilibrium transport process. On the right, the current flow
appears as regions with a decreased voltage compared to the right
lead. This corresponds to an increased chemical potential (excess
electrons due to the transport process).

Small oscillations of the chemical potential with a wavelength on
the order of $\lambda_F / 2$ are apparent in Fig.~\ref{Fig2}c.
They result from interference between paths which emerge from the
leads and directly enter the probe, and paths which first pass the
probe, are reflected from the QPC and only then enter the probe.
This effect was already described in Ref.~\onlinecite{Buttiker2}:
the voltage measurement we make is phase sensitive.
\begin{figure}
\includegraphics[width=8cm]{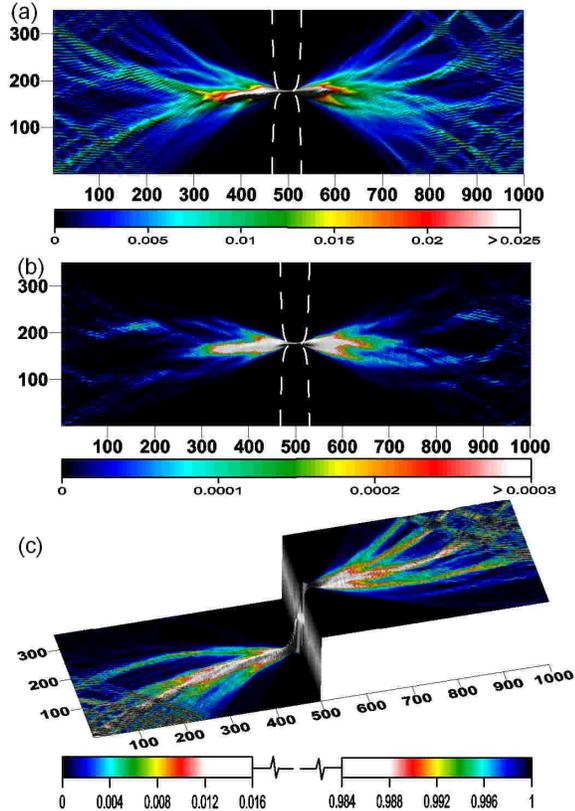}
\caption{\label{Fig3} Maps of electron flow through a quantum
point contact with a moderate magnetic field. Lengths on the axes
are given in units of the lattice parameter $a$. Transport current
density distribution~(a) in units of $2e^2V/(h a)$, STM
conductance map~(b) (conductance is measured in units of $2e^2/h$)
and STM volt probe map~(c). The voltage on the left lead $V_L =
0$, and voltage is measured in units of $V_R$. The Fermi energy
contour is depicted as a dotted white line.}
\end{figure}

In Fig.~\ref{Fig3}, the quantities are calculated for the same
system as before, but now a magnetic field is included. The field
is characterized by a magnetic length of $l_B = 28 \, a$, and a
cyclotron radius $r_c = 835 \, a$. From the non-equilibrium
transport current density plot (which is symmetric in the magnetic
field by its definition in Eqs.~\ref{eqcurrentneq}), it is clear
that the branches of electron flow start bending. The radius of
curvature has the same order of magnitude as the cyclotron radius,
so we are seeing here the onset of the skipping orbit movement of
the electrons. The branches are reflected on the upper and lower
edges of the sample, a proof that one is still in the ballistic
regime.

The conductance difference map (Fig.~\ref{Fig3}b) is quite
unclear. This can be explained as resulting from the reduction of
backscattering in the presence of a magnetic field~\cite{Butt}.
But nevertheless the tendency of the branches to curve can be
observed.

\begin{figure}
\includegraphics[width=8cm]{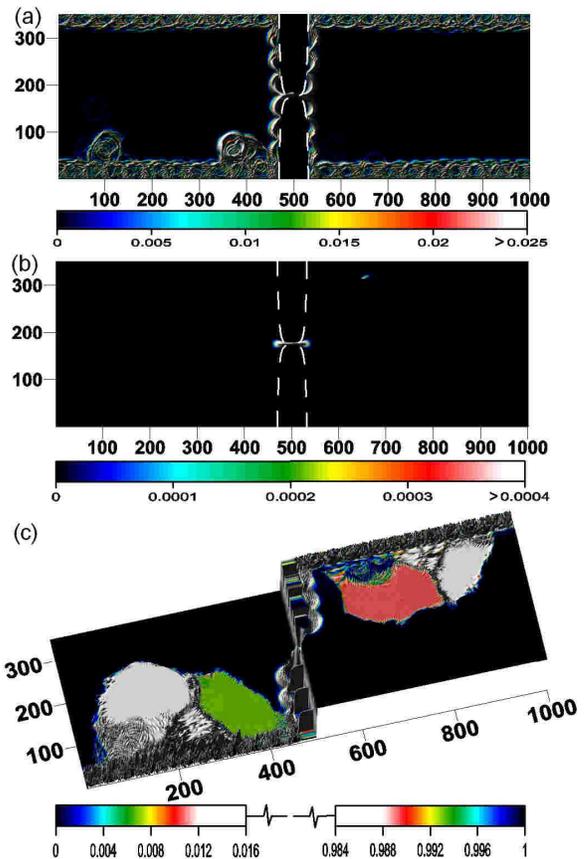}
\caption{\label{Fig4} Maps of electron flow through a quantum
point contact with a high magnetic field. Lengths on the axes are
given in units of the lattice parameter $a$. Transport current
density distribution~(a) in units of $2e^2V/(h a)$, STM
conductance map~(b) (conductance is measured in units of $2e^2/h$)
and STM volt probe map~(c). The voltage on the left lead $V_L =
0$, and voltage is measured in units of $V_R$. The Fermi energy
contour is depicted as a dotted white line.}
\end{figure}

In this regime, the voltage probe method gives better results. The
curved branches are clearly visible in Fig.~\ref{Fig3}c. Please
keep in mind that the voltage method is not symmetric under
reversal of the field, which results in the asymmetry of the
voltage map. This asymmetry will be explained in more detail with
the help of Fig.~\ref{Fig4}c.

In Fig.~\ref{Fig4}, results are shown for a high magnetic field
(magnetic length $l_B = 4.8 \, a$, cyclotron radius $r_C = 24 \,
a$). In a plot of the transport current density the electrons are
seen to describe skipping orbits along the edges of the sample; we
are in the quantum Hall regime. In this regime the original tip
scattering method fails because of a lack of
backscattering~\cite{Butt}. This is visible in Fig.~\ref{Fig4}b:
only in the middle of the quantum point contact there is some
conductance decrease because in this region waves travelling in
opposite directions are ``forced'' to overlap. The map of the
local chemical potential (Fig.~\ref{Fig4}c) gives better results:
the skipping orbits are clearly visible.

The asymmetry of this plot was already pointed out above, and can
be understood as follows: the voltage on the right lead is chosen
to be higher than that on the left lead, so electrons are flowing
from the left to the right. The magnetic field for this plot is
pointing out of the plane of the paper, so electrons emerging from
the left lead flow along the upper left edge of the sample, and
this edge is in equilibrium with the left lead (no skipping orbits
are seen on this edge, only a uniform potential distribution).
Some of these electrons are transmitted through the point contact,
which results in a higher chemical potential (so lower voltage!)
than $\mu_R$ at the upper right edge of the sample. The electrons
reflected from the contact (continuing their path on the lower
edge) give rise to a chemical potential that is lower (= voltage
that is higher) than $\mu_L$ ($V_L$) on the lower left edge.

\subsection{Two Quantum Point Contacts}
\begin{figure}
\includegraphics[width=8cm]{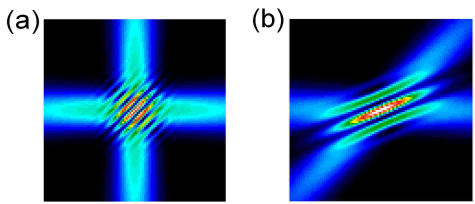}
\caption{ \label{Fig5} Interference between two crossing gaussian
beams. The wavelength of the interference pattern depends on the
angle between the beams. Fringes are more closely spaced for an
angle of $\pi / 2$~(a) than for an angle $\pi / 4$~(b).}
\end{figure}

Looking back at the Figs.~\ref{Fig2} and~\ref{Fig3}, another
interesting interference effect is taking place, which has not
been observed in the experiment. When the branches of electron
flow hit the upper and lower border of the sample (in the regions
from $0$ to $200 \, a$ and from $800 \, a$ to $1000 a$ on both
figures), there are clear interference fringes visible,
perpendicular to the border. The wavelength of these fringes is
larger than that of the fringes observed in the scatterer
experiment (which resulted from back- and forth scattering between
the tip and the QPC). This interference pattern can be explained
as a crossing of two or more coherent electron beams (branches).
In Fig.~\ref{Fig5}, a simulation is shown where the current
density due to two crossing gaussian beams with wavevectors
$\mathbf{k_1}$ and $\mathbf{k_2}$ is calculated. A clear
interference pattern is visible, extending in the direction
$\mathbf{k_1} - \mathbf{k_2}$. From comparison between
Figs.~\ref{Fig5}a and~\ref{Fig5}b, it is clear that the wavelength
of the fringes depends on the angle between the two beams. It can
be shown that this wavelength is given by:
\begin{equation} \label{eqitfwavelength}
\Lambda = \frac{\lambda_F}{2 \sin(\theta / 2)},
\end{equation}
with $ \| \mathbf{k_1} \| = \| \mathbf{k_2} \| = 2 \pi / \lambda_F
$, and $\theta$ the angle between the beams.

\begin{figure}
\includegraphics[width=8cm]{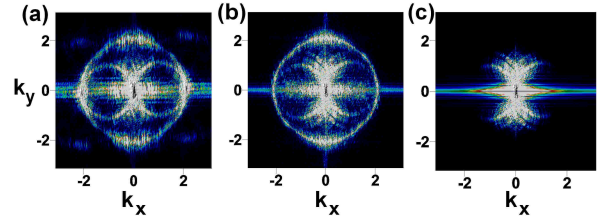}
\caption{ \label{Fig6} Fourier transforms of the flow maps without
a magnetic field: local scatterer method~(a), voltage probe
method~(b), and current density distribution~(c). Wavevectors are
in units of $1/a$, the color scale has arbitrary units.}
\end{figure}

The different periodicities in the different flow maps can be made
more visible by doing a Fourier transformation. In
Fig.~\ref{Fig6}, this is done for columns 1 to 400 of the flow
maps in Fig.~\ref{Fig2} without a magnetic field. In both maps
where an STM is used, a circle with radius $\approx 2 \pi / 3 a$
centered on $(k_x, k_y) = (0,0)$ can be seen. This circle
corresponds to the interference pattern resulting from a
superposition of paths between the STM tip and the quantum point
contact, which creates the fringes spaced at half the Fermi
wavelength. In the current density distribution
(Fig.~\ref{Fig6}c), this circle is of course absent.

Another feature in Fig.~\ref{Fig6} is the presence of two smaller
circles centered on the X-axis. They result from the interference
effect between crossing beams explained above. Using that the
interference pattern of two coherent beams is directed along
$\mathbf{k_1} - \mathbf{k_2}$, together with
Eq.~\ref{eqitfwavelength}, these circles can indeed be reproduced
by having interference between a main beam directed along the
X-axis and others crossing it. While this effect has nothing to do
with scattering off the STM tip, these circles are visible in
Fourier transforms of all flow maps, including that of the current
density distribution.

Now, at crossings between two incoherent beams, an interference
pattern like in Fig.~\ref{Fig5} will not occur. If one looks at
the plot of the transport current with a magnetic field
(Fig.~\ref{Fig3}), one can see that some branches bend upwards,
while other bend downwards under influence of the magnetic field.
This can be interpreted as follows. In the device, the chemical
potential will be somewhere between that of the left and the right
lead. If one assumes that the chemical potential on the left lead
is larger than that on the right, we have an excess of electrons
flowing from left to right. On the other hand we have a deficit of
electrons (``holes'') flowing the other way. These electrons and
holes bend in opposite ways under influence of a magnetic field,
because they fill different scattering states. It is clear then
that electrons and holes, and thus branches curving upwards and
downwards, are emerging from different reservoirs and are thus
phase incoherent. As a result, one does not expect to see
interference between beams with different chirality.
\begin{figure}
\includegraphics[width=8cm]{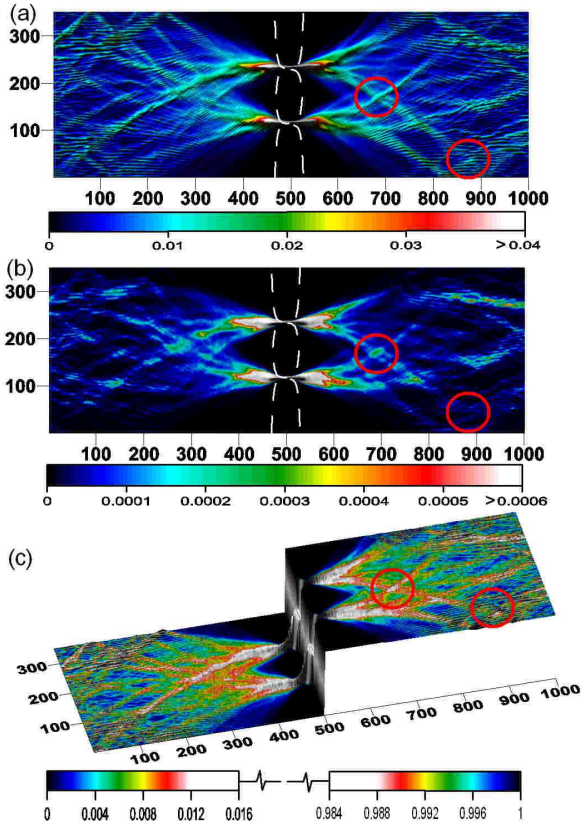}
\caption{ \label{Fig7} Maps of electron flow through a double
quantum point contact with a moderate magnetic field. Lengths on
the axes are given in units of the lattice parameter $a$.
Transport current density distribution~(a) in units of $2e^2V/(h
a)$, STM conductance map~(b) (conductance is measured in units of
$2e^2/h$) and STM volt probe map~(c). The voltage on the left lead
$V_L = 0$, and voltage is measured in units of $V_R$. The voltage
map is symmetrized with respect to the direction of the magnetic
field. In the conductance map, interference fringes with
wavelength $\lambda_F / 2$ resulting from scattering between STM
tip and QPC are smoothed out. The Fermi energy contour is depicted
as a dotted white line.}
\end{figure}

In order to test this statement, we have made some calculations
for a system of two QPC's placed above each other. The potential
of both QPC's has the same form as in Eq.~\ref{eqQPCpotential},
and all parameters are chosen as in Fig.~\ref{Fig3}. The results
are shown in Fig.~\ref{Fig7}. Under influence of the magnetic
field, the branches are curved. One can clearly see the
interference between beams with the same chirality, but there is
no interference at the crossing of two beams with opposite
chirality, like we expected. This distinction becomes clear when
comparing the crossings that are encircled in Fig.~\ref{Fig7}. To
make things more clear, we have smoothed out the interference
fringes in Fig.~\ref{Fig7}b) resulting from scattering between tip
and QPC which had a wavelength of $\lambda_F/2$. In
Fig.~\ref{Fig7}c, we symmetrized the voltage probe plot with
respect to the direction of the magnetic field. Also in this plot,
the behavior for coherent beams crossing is different from that
for incoherent branches. This proves that the effect could be
studied experimentally.

\section{Conclusion}
We have established an efficient tight-binding method to
numerically calculate spatial maps of electron flow as obtained in
a recent series of scanning probe experiments where an STM tip is
used as a local scatterer for electrons~\cite{TopinkaScience}. The
computational effort of our numerical approach scales like $M^3 N$
(in the limit $N \gg 1$), where $M$ is the width of the lattice
and $N$ is the length. It is in this way more efficient than the
standard recursive Green's function method which scales like $M^4
N^2$ for the same problem.

We have also shown expressions for the local density of states,
the electron density and the current density distribution. These
quantities cannot be calculated within the standard recursive
approach, but within our scheme they can be expressed in terms of
the same Green's functions already known from the numerical
simulation of the scanning probe experiment. The computational
effort for these quantities also scales as $M^3 N$.

When a magnetic field is applied, backscattering of electrons will
be strongly reduced because of the presence of edge states so that
the original STM method does not give the desired results.
Therefore, a probe method was proposed where the tip is used to
measure the local chemical potential. For this problem again, the
numerical effort scales like $M^3 N$. The image one obtains from
such a method is not always directly related to the current
pattern in the sample, but one can expect to obtain relevant
information about transport in the sample.

The power of the method was proven in example calculations, where
a tight-binding lattice with more than $350000$ sites has been
used. Moreover, by direct comparison between a numerical
simulation of the experiment and a calculation of the exact
current density distribution, it became clear that the original
scanning probe technique of Topinka et al.~\cite{TopinkaScience}
is really imaging current flow. Furthermore, a new interference
phenomenon has been observed which resulted from the crossing of
phase coherent branches, and a new setup with two QPC's has been
discussed to distinguish between crossings of coherent branches
and incoherent ones. This distinction is visible both when the tip
is used as a scatterer, and when it is used as a voltage probe, so
that an experimental investigation of the effect should be
possible.

It should be clear that the method proposed is very general, and
the information obtained by the different imaging tools very
broad, so that it can be used to study electron flow in a variety
of systems ranging from e.g. the quantum Hall effect to quantum
chaos in electron billiards~\cite{Marcus}. Moreover, including the
spin degrees of freedom proves to be rather easy; every matrix
element should be replaced by a spinor. As such, an even broader
range of phenomena could be studied, ultimately also those
including spin-orbit coupling (e.g. the spin Hall
effect~\cite{Hirsch, Sinova}).

\appendix

\begin{figure*}
\includegraphics{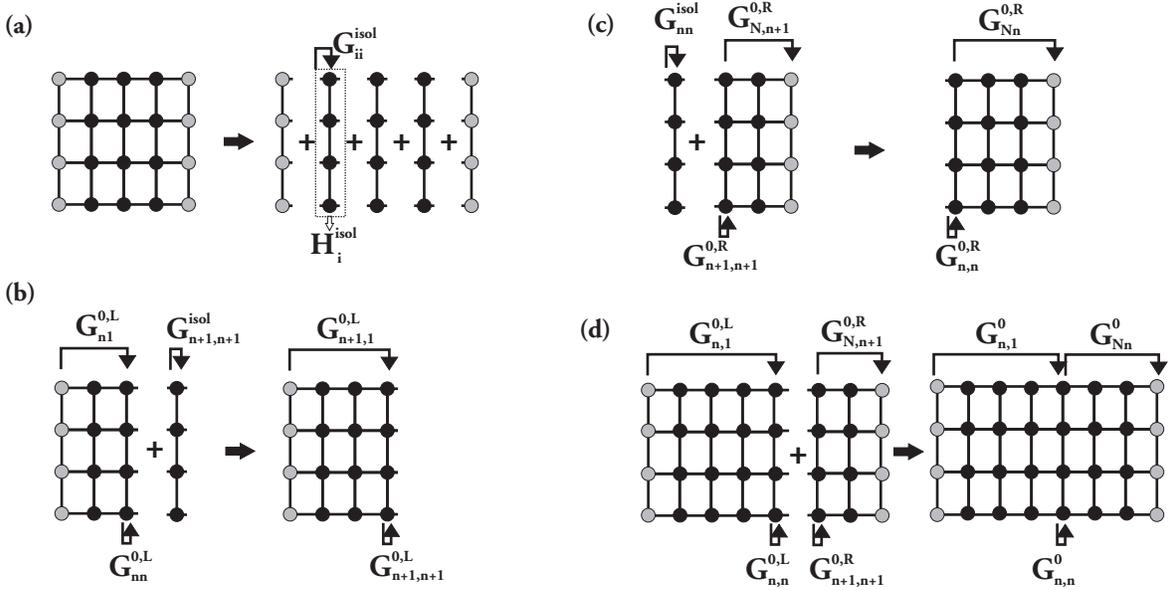}
\caption{\label{Fig8} Green's function method. After the hopping
matrices between columns are set to zero to separate the device in
vertical slices, and their Green's functions
$G_{ii}^{\text{isol}}$ are calculated~(a), the slices are added
one by one to obtain $G_{N1}^{0,L}$~(b). Then, the isolated slices
are added to each other again but now starting from the right~(c).
Finally, by combining sections pairwise, one can obtain the
$G_{Nn}^{0}$, $G_{nn}^{0}$ and $G_{n1}^{0}$~(d). In the figure,
the first and last column of the device are colored grey to denote
the influence of the self-energies of the leads.}
\end{figure*}

\section{Calculation of Green's Functions relevant to our Problem}
\label{AppendixA} In the paper, it became clear that indeed all
quantities we need can be expressed in terms of a small subset of
Green's functions $G_{n1}^{0}$, $G_{nn}^{0}$, $G_{nN}^{0}$ and
$G_{Nn}^{0}$ (see Fig.~\ref{Fig1}). In this appendix, we will
treat in detail how to obtain these functions with a numerical
effort that scales like $M^3N$.

The first step is to divide the central device into its separate
columns, and to put the hopping matrices between them equal to
zero so that the columns become isolated, as depicted in
Fig~\ref{Fig8}a. Next, one calculates the Green's function
$G_{ii}^{\text{isol}}$ for every isolated column $i=1,2,...,N$,
which amounts to doing a single $M \times M$ matrix inversion.
This first step thus needs a total of $N$ inversions.

The next step is to attach the isolated columns one by one to each
other by including the hopping matrices between them using Dyson's
equation (this is like in the standard recursive technique, see
e.g. Ref.~\onlinecite{Ferry}). As such, one calculates the Green's
functions $G_{n1}^{0,L}$ and $G_{nn}^{0,L}$ (see
Fig.~\ref{Fig8}b). Please note that since not all columns are
attached to each other in $G_{n1}^{0,L}$ and $G_{nn}^{0,L}$ (only
those to the left of column $n$), these Green's functions are not
equal to the Green's functions $G_{n1}^{0}$ and $G_{nn}^{0}$ we
are trying to obtain (an exception of course is $G_{N1}^{0,L}$
which in fact is equal to $G_{N1}^{0}$). The superscript L is
added to make this distinction clear.

Attachment of a single isolated column costs one inversion of an
$M \times M$ matrix, so a total of $N-1$ matrix inversions are
necessary to calculate the $G_{n1}^{0,L}$ and $G_{nn}^{0, L}$ for
all $n$, starting from the $G_{ii}^{\text{isol}}$.

For the third step, we start over from the isolated Green's
functions already calculated in step 1, and glue them together
like we did in the previous step on the basis of a Dyson's
equation, but now beginning from the right, like depicted in
Fig.~\ref{Fig8}c. The Green's functions we calculate with every
step are $G_{Nn}^{0,R}$, $G_{nn}^{0,R}$ and $G_{nN}^{0,R}$. Again,
a single matrix inversion is needed for the attachment of a single
column, so a total of $N-1$ inversions for the completion of the
third step.

The final step is to get the $G_{Nn}^{0}$, $G_{nN}^{0}$,
$G_{nn}^{0}$ and $G_{n1}^{0}$ we are looking for by attaching the
previously calculated Green's functions in pairs, like illustrated
in Fig.~\ref{Fig8}d. One takes a section of connected columns $1$
to $n$ (with known Green's functions $G_{n1}^{0,L}$ and
$G_{nn}^{0,L}$), and attaches it to the section of connected
columns $n+1$ to $N$ (with Green's functions $G_{N,n+1}^{0,R}$ and
$G_{n+1,n+1}^{0,R}$). Please note that a single inversion is
necessary for each pairwise addition, which gives a total of $N-1$
inversions for the final step.

From the previous discussion, it becomes clear that we need $N +
3(N-1)$ inversions for calculating $G_{Nn}^{0}$, $G_{nN}^{0}$,
$G_{nn}^{0}$ and $G_{n1}^{0}$ for all $n$. So our method scales
indeed linear with $N$ in the number of inversions.

\section{Evaluation of the Trace} \label{AppendixB}
For the scatterer method, it is necessary to calculate the
conductance difference in Eq.~\ref{eqtransw}. For this, it seems
that we have to evaluate the trace in Eq.~\ref{eqtransmission2}
for all $MN$ tip positions. This trace contains products of $M
\times M$ matrices, so the numerical effort for this step would
scale like $M^4 N$. As such one would loose a factor of $M$ in
efficiency compared to the rest of the calculation. However, there
is a better way to evaluate the conductance difference in
Eq.~\ref{eqtransw}.

We write (see Eq.~(\ref{eqDysonperturb})):
\begin{equation} \label{eqA}
G_{N1} = G_{N1}^0 + A,
\end{equation}
with the $M \times M$ matrix:
\begin{equation}
A = G_{Nn}^{0} \, V_{nn}^{\text{tip}}\, (1 - G_{nn}^{0} \,
V_{nn}^{\text{tip}})^{-1} \, G_{n1}^{0}.
\end{equation}
It is important now to make it clear that since
$V_{nn}^{\text{tip}}$ has only one non-zero element, namely on
position $(m,m)$, one can write $A$ as a product of a column
matrix and a row matrix:
\begin{equation}
A = [G_{Nn}^0]_{m^\text{{th}} \text{column}} \  \tau \
[G_{n1}^0]_{m^\text{{th}} \text{row}},
\end{equation}
with the scalar $\tau$ given by ($v$ is the magnitude of the
repulsive tip potential):
\begin{equation}
\tau = \frac{v}{1-v \, \langle m | G_{nn}^{0} | m \rangle}.
\end{equation}
By substituting Eq.~(\ref{eqA}) into Eq.~(\ref{eqtransmission2}),
one obtains:
\begin{equation} \label{eqtranslast}
T(m,n) = T^0 + 2 \text{Re} \, \text{Tr} \, [\Gamma_R A \Gamma_L
(G_{N1}^0)^{\dag}] + \text{Tr} [\Gamma_R A \Gamma_L A^{\dag}],
\end{equation}
So in order to evaluate the conductance difference $\Delta g
(m,n)$, we need to evaluate only the last two terms in
Eq.~(\ref{eqtranslast}). The last term only involves products of
an $M \times M$ matrix with row or column matrices because of the
special form of $A$. The computational effort for this term scales
thus as $M^2$, which corresponds to a total effort of $M^3 N$ for
all tip locations. Now, since $\Gamma_L (G_{N1}^0)^{\dag}$ is
independent of the tip position, it has to be calculated only once
(with an effort $M^3$). When this matrix is known, the trace in
the second term also contains only products of an $M \times M$
matrix with a row or column matrix, so the computational effort
then scales like $M^2$, and as a result the effort for all tip
positions scales like $M^3 N$ in the limit of large $N$. As such,
we do not loose our efficiency.

\end{document}